\documentclass[aps,prd,twocolumn,superscriptaddress,preprintnumbers,floatfix,nofootinbib,notitlepage,showkeys,showpacs]{revtex4-1}

\usepackage[utf8]{inputenc}

\usepackage{graphicx}
\usepackage{hyperref}
\usepackage{amsmath}
\usepackage{amssymb}
\usepackage{color}

\newcommand{\be}{\begin{equation}} 
\newcommand{\ee}{\end{equation}}
\newcommand{\bea}{\begin{equation}\begin{aligned}}
\newcommand{\eea}{\end{aligned}\end{equation}}
\newcommand{\td}{{\rm d}}

\def\lsim{\mathrel{\raise.3ex\hbox{$<$\kern-.75em\lower1ex\hbox{$\sim$}}}}
\def\gsim{\mathrel{\raise.3ex\hbox{$>$\kern-.75em\lower1ex\hbox{$\sim$}}}}

\newcommand{\yr}{{\rm yr}}
\newcommand{\kyr}{{\rm kyr}}
\newcommand{\Myr}{{\rm Myr}}

\newcommand{\Gpc}{{\rm Gpc}}

\newcommand{\Msun}{M_\odot}

\begin{document}

\preprint{CERN-TH-2019-141, KCL-PH-TH/2019-69}

\title{Lower bound on the primordial black hole merger rate}

\author{Ville Vaskonen}
\email{ville.vaskonen@kcl.ac.uk}
\affiliation{Physics Department, King's College London, London WC2R 2LS, UK}
\affiliation{NICPB, R\"{a}vala 10, 10143 Tallinn, Estonia}
\author{Hardi Veerm\"ae}
\email{hardi.veermae@cern.ch}
\affiliation{NICPB, R\"{a}vala 10, 10143 Tallinn, Estonia}
\affiliation{Theoretical Physics Department, CERN, CH-1211 Geneva 23, Switzerland}

\begin{abstract}
We derive a lower bound on the merger rate of primordial black hole (PBH) binaries by estimating the maximal fraction of binaries that were perturbed between formation in the early Universe and merger, and computing a conservative merger rate of perturbed binaries. This implies robust constraints on the PBH abundance in the range $1-100 M_\odot$. We further show that LIGO/Virgo design sensitivity has the potential to reach the PBH mass range of $10^{-2}-10^3 M_\odot$. The constraint from the merger rate of perturbed binaries is stronger if PBHs are initially spatially clustered.
\end{abstract}

\maketitle

\section{Introduction}

The direct detection of gravitational waves (GWs) from the binary black hole (BH) merger by LIGO~\cite{Abbott:2016blz} marked the dawn of a new era in cosmology. During the first two observing runs LIGO detected nine more BH-BH merger events, latest of which were seen also by the Virgo detector, indicating a merger rate of $9.7 - 101 \Gpc^{-3} \yr^{-1}$ for BH binaries with component masses ranging from $7.6M_\odot$ to $50.5M_\odot$~\cite{LIGOScientific:2018mvr}. Currently possible new events are reported almost on a weekly basis~\cite{GraceDB}. These events may have originated from astrophysical BHs~\cite{Belczynski:2016obo} or from primordial black holes (PBHs)~\cite{Bird:2016dcv,Clesse:2016vqa,Sasaki:2016jop} that, in the most common scenarios, arise from large curvature fluctuations in the early Universe~\cite{Hawking:1971aa,Carr:1974nx}.

PBH binary formation has received much attention in the last years~\cite{Bird:2016dcv,Clesse:2016vqa,Sasaki:2016jop,Clesse:2016ajp,Wang:2016ana,Wang:2019kaf,Raidal:2017mfl,Ali-Haimoud:2017rtz,Ballesteros:2018swv,Kavanagh:2018ggo,Raidal:2018bbj,Garriga:2019vqu}. Most PBH binaries whose mergers would be presently observed form already in the radiation dominated Universe~\cite{Sasaki:2016jop,Raidal:2017mfl,Ali-Haimoud:2017rtz}. Due to the random distribution of PBHs before formation of structures, some PBHs may be much closer than their average separation. Such PBHs will form the earliest gravitationally bound systems and produce the initial population of binaries. The distribution of orbital characteristics of this binary population can be estimated analytically and, assuming all binaries remain unperturbed until merger, the merger rate obtained via this mechanism is so large that only a small fraction, $\mathcal{O}(0.1\%)$, of dark matter (DM) in PBHs is allowed in the mass range that LIGO/Virgo detectors probe. This implies the strongest constraints on the PBH abundance to date in the mass range $1-100\Msun$~\cite{Raidal:2017mfl,Ali-Haimoud:2017rtz,Raidal:2018bbj}. Yet, the fate of these binaries remains uncertain. In particular, initial binaries are highly eccentric, thus interactions with surrounding PBHs can significantly increase their coalescence times. 

The aim of this paper is to derive a lower bound on the PBH merger rate by considering scenarios where the initial population of binaries is maximally perturbed and by including the contribution from perturbed binaries. These results imply more reliable constraints on the PBH abundance from GW observations.

Numerical studies, although confirming the analytical orbital parameter distribution of initial binaries, find that binaries are likely to be perturbed after formation in the case when PBHs make up most of the DM~\cite{Raidal:2018bbj}.\footnote{We consider disruption (or perturbation) of binaries in a very broad sense: This includes modification of the orbital parameters due to tidal torque, hard collisions with other BH resulting in a binary (including the ones where a BH in initial the binary is swapped), and ionization of the binaries. For hard binaries, only the first two scenarios are probable. } The binaries can be perturbed by two mechanisms: (1) In case the initial configuration contains a third PBH close to the PBH pair that is expected to form a binary, it is very likely that it collides with the binary. (2) PBHs will form dense $N$-body systems relatively early, and binaries absorbed by these clusters become more likely to be perturbed.  

In~\cite{Raidal:2018bbj} we introduced a suppression factor of the merger rate due to disruption by the first mechanism. In this paper we consider the disruption via the second mechanism. Analytic estimates of Ref.~\cite{Ali-Haimoud:2017rtz} indicate that most of the initial PBH binaries are not perturbed between formation and merger in typical DM haloes larger than 10 PBH, assuming that early haloes are disrupted when absorbed by subsequent larger haloes. We show that the dense haloes forming at redshifts $z \gtrsim 100$ can significantly perturb their initial binary population in case they survive absorption by their later hosts. Furthermore, self-gravitating systems are not stable, and, due to the gravothermal catastrophe~\cite{LyndenBell:1968yw,2008gady.book.....B}, the binary may end up in an environment where the PBH density is several orders of magnitude larger than what was used in Ref.~\cite{Ali-Haimoud:2017rtz}, for example, when a binary finds itself in the central region of a PBH cluster undergoing a core collapse. This increases the rate of close encounters and the probability of perturbing the binary. To obtain the strongest possible suppression for the merger rate of initial binaries we assume that clusters small enough to experience a core collapse within a Hubble time will perturb all of their initial binaries. Larger haloes are stable and do not perturb their initial binaries due to their smaller densities and higher velocity dispersions.

In this paper we also improve our previous estimate~\cite{Raidal:2018bbj} of the merger rate arising from the population of \emph{perturbed} binaries. Even if most of the early PBH binaries were at least once perturbed, the present merger rate from the resulting population of less eccentric binaries can still reach the observed rate. In that case, binaries contributing to the rate originate from dense 3-body configurations with separations much below the average PBH distance. 

By combining the merger rate of perturbed binaries with a maximally suppressed merger rate of initial binaries we derive a lower bound for the merger rate, concluding that it is above the range indicated by the LIGO and Virgo observations in the case that more than 4\% of DM is in $\mathcal{O}(10M_\odot)$ PBHs. We finally use our results to obtain robust constraints on PBH abundance. Throughout this paper we use geometric units $G_{N} = c = 1$.

\section{PBH binaries from the early Universe}

The coalescence time of a binary due to GW emission is approximately~\cite{Peters:1964zz}\footnote{This holds when $j \ll 1$, and for larger $j$ deviates by at most 23\%.}
\be \label{eq:tau}
	\tau 
	= \frac{3}{85} \frac{r_a^4}{\eta M^3} j^7 
	= \frac{3}{1360} \frac{M}{\eta E^4} j^7 
\ee
where $\eta = m_1 m_2/M^2$, $M=m_1+m_2$ denote the mass asymmetry and the total mass of the binary,  $r_a$ is its semimajor axis, $E = M/(2r_a)$ the binding energy per reduced mass, $j \equiv \sqrt{1- e^2}$ the dimensionless angular momentum and $e$ is the eccentricity. 

Binaries expected to coalesce within a Hubble time are hard. Thus, according to the Heggie-Hills law~\cite{1975MNRAS.173..729H,1975AJ.....80..809H}, close encounters with other PBHs will, on average, increase their binding energy by an $\mathcal{O}(1)$ factor. In particular, these binaries are very unlikely to be ionized. On the other hand, the initial binaries are highly eccentric. The characteristic angular momentum for initial binaries with coalescence time $\tau$ is~\cite{Raidal:2018bbj}
\be\label{eq:j_tau}
	j_{\tau} =  0.02 f_{\rm PBH}^\frac{16}{37} (4\eta)^{\frac{3}{37}}\left[\frac{M}{\Msun}\right]^{\frac{5}{37}} \left[\frac{\tau }{t_0}\right]^{\frac{3}{37}},
\ee
where $t_0$ is the age of the Universe. Thus, $j$ very likely increases by more than an order of magnitude leading to an over 7 orders of magnitude increase of the coalescence time. So, when a binary that was initially expected to merge within a Hubble time is perturbed, its coalescence time exceeds the age of the Universe, and thus it will not produce detectable GW signals. We stress that, due to their lower eccentricities, the last effect is insignificant for perturbed binaries.

If the initial population would not be perturbed by the second process, then a fraction $f_{\rm PBH} \geq 10^{-3}$ of PBH DM predicts a merger rate higher than observed by LIGO/Virgo. The differential merger rate at time $t$ in that case is~\cite{Raidal:2018bbj}
\bea\label{eq:R_unperturbed}
	\frac{\td R_{\rm np}}{\td m_1 \td m_2} \approx &\frac{1.6 \times 10^{6}}{\Gpc^{3} \yr} \, f_{\rm PBH}^{\frac{53}{37}}  \left[\frac{t}{t_{0}}\right]^{-\frac{34}{37}} \left[\frac{M}{\Msun}\right]^{-\frac{32}{37}}  \\
	& \times S[\psi,f_{\rm PBH},M] \, \eta^{-\frac{34}{37}} \psi(m_1) \psi(m_2) \,, 
\eea
$\psi(m)$ is the PBH mass distribution, and $S[\psi,f_{\rm PBH}, M]$ is a suppression factor accounting for the disruption by the first mechanism due to the infall of the PBH close to the binary. For narrow mass functions $S \approx 0.24 (1 + 2.3 \sigma_M^2/f_{\rm PBH}^2)^{-21/74}$, where $\sigma_M\simeq 0.005$ is the variance of matter density perturbations at the time the binary was formed.  In the rest of the paper we will consider only monochromatic mass functions, $\psi(m') = \delta(m'-m)$, and denote the PBH mass by $m$.

When $f_{\rm PBH} \gsim 0.1$, the fraction of binaries perturbed by the second mechanism, that is, by PBH clusters, has been shown to be relatively high already at $z = 1100$, indicating that nearly all initial binaries might be perturbed within the age of the Universe~\cite{Raidal:2018bbj}. A small fraction of the early binaries may, however, remain unperturbed. So, by~\eqref{eq:R_unperturbed}, the present merger rate from the initial binary population can still exceed the current bounds in the LIGO/Virgo mass range. 

This may happen because of several reasons: (a) not all PBH binaries will become bound to PBH clusters early; (b) small dense clusters of PBH are unstable on timescales much shorter than the Hubble time, and thus it is possible that the cluster is dissolved before the binary is perturbed; (c) larger, less dense systems forming the DM haloes of, e.g. dwarf galaxies, must be stable within the Hubble time making disruption unlikely.

\section{Disruption of initial PBH binaries in small haloes}

Let us estimate the probability that a binary will be significantly perturbed via encounters with other PBH. The characteristic timescale for this process is given by
\be \label{eq:tpinv}
	1/t_{\rm p} = n_{\rm loc} \, \langle \sigma_{\Delta j > j_{\tau}} v  \rangle \,,
\ee
where $n_{\rm loc}$ is the local PBH number density, $v$ the perturber velocity, and $\sigma_{\Delta j > j_{\tau}}$ the cross-section for increasing the angular momentum of the initial binary by an amount comparable to its initial value \eqref{eq:j_tau}. Note that, by Eq.~\eqref{eq:tau}, this corresponds to a 2 order of magnitude increase in the coalescence time. The average change of angular momentum is roughly $\Delta j \approx m^{1/2} r_{a}^{3/2}/(r_{c} b v)$
, where $b$ is the impact parameter and $r_{c}$ is the distance of closest approach~\cite{Ali-Haimoud:2017rtz}. By conservation of angular momentum and energy, they are related as $b^2 = r_{c}^2 + 6m r_{c}/v^2$. The second term dominates in the early Universe where close encounters are more likely due to the lower velocities and higher densities. In this case
\be\label{eq:sigma_p}
	\sigma_{\Delta j > j_{\tau}} = \pi b^2
	\approx \frac{28 m^{7/4} \tau^{1/4} }{v^2 j_{\tau}^{29/12}} \,.
\ee

Consider now the early haloes containing $N$ PBHs. They form approximately when the scale factor is $a_c \equiv (1+z_c)^{-1} \approx a_{\rm eq} \sqrt{N}/f_{\rm PBH}$, where $a_{\rm eq}$ corresponds to matter-radiation equality. Assuming that they are virialized, the velocity dispersion is given by $\sigma_v^2 \approx M_{\rm H}/R$, where $R$ is the virial radius and $M_{\rm H} = m N/f_{\rm PBH}$ the mass of the halo\footnote{We assume that the fraction of PBHs in haloes matches $f_{\rm PBH}$. However, it was shown in~\cite{Inman:2019wvr} that this fraction could be larger, especially in the early Universe. This may slightly enhance the disruption of PBH binaries.} Following Press-Schechter theory, the average energy density of matter in these haloes is $\rho = 3M_{\rm H}/{4\pi R^3} \approx 18 \pi^2 \rho_{c} a^{-3}$, where $\rho_c$ is the critical comoving density.

Using $v \approx  \sigma_v$ and a density that is $18 \pi^2$ times above the average, we obtain from \eqref{eq:tpinv} and \eqref{eq:sigma_p} that
\be\label{eq:t_p}
	t_{\rm p} 
	= 2 \Myr \times N^\frac13 f_{\rm PBH}^{-\frac{32}{111}} \left[\frac{m}{\Msun}\right]^{-\frac{10}{111}} \left[\frac{1+z_c}{1000}\right]^{-\frac{5}{2}} \,.
\ee
Therefore, when $m = \mathcal{O}(10M_\odot)$ binaries in haloes with $N\lsim 2000$ are perturbed before today if $f_{\rm PBH}=1$. Since, by definition, binaries can be perturbed only in systems with $N_{c} \geq  3$ PBH, binaries are unlikely to be perturbed when $f_{\rm PBH} \lesssim 0.02$. 

These estimates are applicable assuming that the small haloes formed at high redshifts survive, i.e. they maintain their initial density and velocity dispersion. This assumption is, however, easily violated as the haloes can be absorbed into larger structures or expand due to the heating provided by binary-PBH collisions. Both effects reduce the frequency of binary-PBH encounters. In fact, it was estimated that the disruption of initial binaries in DM haloes can be neglected when the earlier smaller haloes are continuously absorbed and disrupted by the subsequent generation of larger haloes~\cite{Ali-Haimoud:2017rtz}.

By drawing parallels with globular clusters~\cite{Sigurdsson:1994ju,Sigurdsson:1993zrm}, there are two effects that can significantly enhance the disruption probability in PBH haloes: First, since binaries are heavier than a single PBH, they tend to sink towards the center of the halo. Second, during core collapse  the central density can increase indefinitely until the collapse is stopped by binary-PBH interactions heating the core.~\cite{2008gady.book.....B}. This will switch on 3-body encounters also in larger haloes in which they were initially unlikely. Thus, to estimate the maximal disruption probability by the second mechanism we assume that all binaries within a cluster that is unstable within a timescale less than a Hubble time are perturbed. 

For the timescale of the gravothermal instability we use the characteristic time of core collapse given by $t_{\rm cc} \geq 18 t_{\rm r}$, where the relaxation time is~\cite{Quinlan:1996bw}
\be
	t_{\rm r} = 0.065 \frac{\sigma_v^3}{m \rho \ln \Lambda} 
	\approx 2 \kyr \frac{N^{7/4}}{f_{\rm PBH}^{5/2} \ln \Lambda} \,.
\ee
The Coulomb logarithm is approximately $\ln\Lambda \approx \ln (N/f_{\rm PBH})$. Requiring $18t_{\rm r} < t_0$ gives
\be\label{eq:Nc}
	N \leq N_c \equiv 1500 f_{\rm PBH}^{10/7} \ln\Lambda^{4/7} \,,
\ee
which is consistent with earlier results of~\cite{Afshordi:2003zb}. We now find that binaries are perturbed in haloes with $N<5300$ if $f_{\rm PBH}=1$ and all binaries survive if $f_{\rm PBH} \lesssim 0.005$. We note that binary-PBH collisions can heat the halo and stop the collapse; thus, disruption of initial binaries due to core collapse is relevant for haloes containing $2000 \lsim N \lsim 5300$ PBH when $f_{\rm PBH} \approx 1$.

Finding the probability for an initial PBH binary to be disrupted thus boils down to finding the fraction of initial binaries in unstable haloes. The distribution of haloes containing $N$ PBHs at redshift $z$ from initially Poisson distributed point masses is approximately~\cite{1983MNRAS.205..207E,Hutsi:2019hlw,Inman:2019wvr}
\be \label{eq:halomf}
	p_{N}(z) \propto N^{-1/2}e^{-N/N^{*}(z)} \,,
\ee
where $N^{*}(z)$ is the characteristic number of PBH in a halo at redshift $z$, which we estimate using analytic results from~\cite{Inman:2019wvr}. These haloes can have substructure which, following~\cite{Gao:2004au}, we assume to be distributed by the halo mass function~\eqref{eq:halomf}. The probability of finding a binary in a halo with $N$ PBHs is approximately proportional to $p_N$\footnote{We remark, that binaries for which the surrounding PBH density was initially larger have, on average, a higher initial angular momentum and must thus have a smaller initial separation if they are to merge within a Hubble time. Therefore, there are fewer initial binaries in dense haloes. Such haloes are likely small.} and the probability of finding a PBH in a subhalo of $N$ PBHs inside a halo of $N'>N$ PBHs to $p_N p_{N'}$. So, the fraction of nonperturbed binaries is bounded below by
\bea \label{eq:pnp}
	P_{\rm np}(z) \gtrsim 1 - &\sum_{N=3}^{N_{c}(z)} \bar p_{N}(z_c) \\& - \sum_{N'>N_{c}(z)} \left[\sum_{N=3}^{N_{c}(z)} \tilde p_{N}(z_c)\right] \bar p_{N'}(z_c) \,,
\eea
where $N_{c}(z)$ defined in Eq.~\eqref{eq:Nc} is the smallest number of PBH in haloes or subhaloes that are expected to be stable until redshift $z$, and the halo distribution is evaluated at the redshift $z_c$ at which the haloes with $N_{c}$ PBHs formed. The probabilities $\bar p_N$ and $\tilde p_N$ are normalized as
\be
	\sum_{N \geq 2} \bar p_N = 1 \,, \qquad
	\sum_{N = 2}^{N'} \tilde p_{N} = 1 \,. 
\ee
Here we assume that all subhaloes survive and that every PBH inside a halo belongs to some subhalo. As substructure can be absorbed or disrupted by the host, this construction will overestimate the probability for the initial binary to be perturbed leading to a conservative merger rate for initial binaries.

\begin{figure}
\centering
\includegraphics[width=0.39\textwidth]{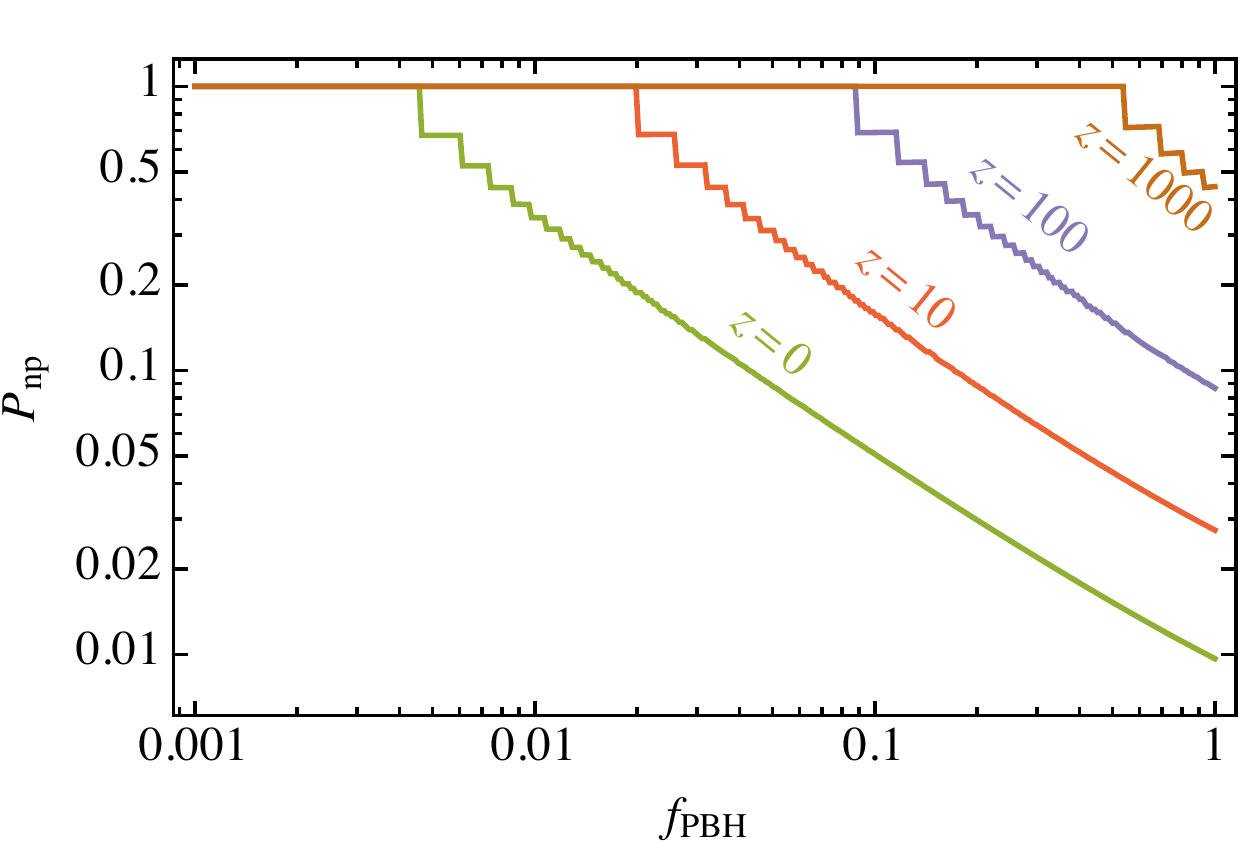}
\caption{Lower bound on the suppression factor of the merger rate of initial binaries as a function of PBH abundance shown at different redshifts. It is estimated as the fraction of PBHs that reside in (sub)haloes that are gravitationally stable until the given redshift.}
\label{fig:pnp}
\end{figure}

In Fig.~\ref{fig:pnp} we show the strongest suppression factor $P_{\rm np}(z)$ of the merger rate. For $f_{\rm PBH}=1$ at most $6\%$ of the initial binaries survive unperturbed until today. Moreover, disruption by clusters can be observed at very high redshifts. This is consistent with our $N$-body simulations~\cite{Raidal:2018bbj}, according to which only about half the binaries not perturbed by the first mechanism survive until $z\simeq 1100$ when $f_{\rm PBH}=1$.

\section{Merger rate of perturbed binaries}

Even if all early binaries would be perturbed, a large population of PBH binaries capable of contributing to the present merger rate still remains. In the following we will estimate the merger rate of perturbed binaries. 

Consider first the initial population of binaries and the resulting merger rate. Following~\cite{Raidal:2018bbj} we approximate that all initial pairs are perturbed when there is a third PBH closer than $y_{\rm min}$, determined by $N(y_{\rm min}) = 2$, where $N(y) \equiv 4\pi n_{\rm PBH} y^3/3$ is the expected number of PBH in the comoving volume sphere of radius $y$ and $n_{\rm PBH}$ is the PBH number density. This comprises $76\%$ of all initial PBH binaries (for $f_{\rm PBH}\gsim 1.5 \sigma_M \approx 0.01$) and the rate~\eqref{eq:R_unperturbed} results from the remaining $24\%$. We stress that the initial binary may be perturbed even if the third PBH is farther than $y_{\rm min}$ by the mechanism discussed in the previous section.

Most of the perturbed PBH binaries whose mergers can be observed today would come from PBH pairs that initially formed an eccentric binary with a very short coalescence time. It is thus necessary to impose that the initial coalescence time~\eqref{eq:tau} is larger than the time required to perturb the binary, $\tau_{\rm i} \gg t_{\rm p}$. The first close encounter of the initial pair takes place at $a \approx a_{\rm eq} N(x)/(2 f_{\rm PBH})$, where $x$ denotes the initial comoving separation of the pair. Analogously, the initial encounter of the binary with the third PBH takes place at $a = a(t_p) \approx a_{\rm eq} N(y)/(3 f_{\rm PBH})$. This estimate works well when the 3-body system is only weakly coupled to the surrounding PBHs, implying $N(y) \lesssim 1$. For larger values of $y$, it is more likely that the binary is not the closest object to the third PBH. If the 3-body system forms during radiation domination, we obtain 
\be \label{eq:t_p}
	t_{\rm p}(y) \approx 7 \, \kyr \times N(y)^2 f_{\rm PBH}^{-2} \,.
\ee
For short coalescence times GW emission may be relevant. The energy of eccentric orbits scales as $E(t) = E_0 (1-t/\tau)^{-2}$~\cite{Peters:1964zz}. Thus, to avoid binaries that emit more than 10\% of their initial binding energy in GWs, we consider a population in which the coalescence time of the initial binary satisfies
\be\label{eq:taubound}
	\tau_{\rm i} > 20 t_{\rm p}(y) \,.
\ee
In the matter dominated epoch~\eqref{eq:t_p} overestimates $t_{\rm p}$, and therefore requiring~\eqref{eq:taubound} leads to a lower merger rate estimate.

The initial coalescence time depends on the initial angular momentum which, assuming a hierarchical collapse, is determined by the tidal forces acting on the pair.  The closest PBH generates an angular momentum $j_1 = 0.7  |\sin(2\theta)|N(x) /N(y)$, where $\theta$ is the angle between the vector joining the PBH pair and the vector joining the center of mass of the pair and the position of the third PBH. The angular momentum generated by tidal forces from all other surrounding PBHs is of the order $j_0 \approx 0.5 N(x)$, where the width of the distribution is $\sigma_{j} \approx 0.5 N(x) /\sqrt{N(y)}$~\cite{Raidal:2018bbj}. If $N(y) \ll 1$, then $j_1 \gtrsim \sigma_{j}  \gg j_0$ as long as $|\sin(2\theta)| \gtrsim \sqrt{N(y)}$. In this case the contribution from other PBHs is negligible and we can use the 3-body approximation.

The comoving density of binaries with an initial binding energy in the interval $(E,E+\td E)$ is 
\be \label{eq:niE0}
	\frac{\partial n_i(E)}{\partial E} = \int \td n_{\rm pairs} \, P(E,\tau_i \geq 20 t_p|x,y,\theta) \,,  
\ee
where
\be \label{eq:npairs}
	\td n_{\rm pairs} = \frac{n_{\rm PBH}}{2}\td N(x) \td N(y) e^{-N(y)} \frac{\td \cos \theta}{2}
\ee
is the density of initial configurations specified by $x$, $y$ and $\theta$, and 
\bea \label{eq:Pconf}
 	P(&E,\tau_i \geq 20 t_p|x,y) \\
	&=\delta(E - E(x))\Theta(\tau_{\rm i}(E,j_1(x,y,\theta)) - 20 t_{\rm p}(y))
\eea
is the probability that such initial configurations produce a binary with binding energy $E$ satisfying~\eqref{eq:taubound}. $\Theta$ denotes the step function. The relation between binding energy and initial separation is approximately $x \approx 1.2 \sqrt{m}(E(x) \rho_R)^{-1/4}$, where $\rho_R$ is the comoving radiation energy density, and the coalescence time is given by~\eqref{eq:tau} with $j = j_1(x,y,\theta)$. Requiring that the distance to the third PBH always exceeds the separation of the pair implies $y > 3/2x$, i.e. $N(y)> 27/8 N(x)$.  For consistency, we exclude initial conditions with $N(y)>1$. The condition~\eqref{eq:taubound} can be recast as $N(y) < 0.36 N(x) |\sin(2\theta)|^{\frac{7}{9}} E^{-\frac{5}{18}}$, and we further require that the r.h.s is smaller than one at all angles. This implies a lower bound on $E$,
\be
	E > \left[15 {\rm \frac{km}{s}}\right]^2  \left[\frac{m}{\Msun}\right]^{\frac{18}{37}} f_{\rm PBH}^{\frac{36}{37}} \,,
\ee 
that guarantees the hardness of the binaries. At the relevant binding energies the distribution \eqref{eq:niE0} scales as $\partial n_i(E)/\partial E \propto E^{-\frac{25}{9}}$;  it is independent of the mass of the PBHs and proportional to $f_{\rm PBH}^3$.

After the initial binaries have interacted with the surrounding PBHs their orbital parameter distribution is approximately
\be \label{eq:evo_n}
	\frac{\partial^2 n_{p}(j,E)}{\partial j \partial E} = \frac{\partial P_{p}(j)}{\partial j} \int \td E' \frac{\partial K (E | E')}{\partial E} \frac{\partial n_{i}(E')}{\partial E'} \,, 
\ee
where $\partial P_{p}(j)/\partial j$ is the angular momentum distribution of the perturbed binaries and $K(E | E')$ is the energy distribution of perturbed binaries with initial energy $E'$. The assumption that the angular momentum distribution $P(j)$ of perturbed binaries is independent of the initial angular momentum agrees with numerical studies of binary-single PBH collisions~\cite{1993ApJ...415..631S,Fregeau:2004if}. 

 The thermal distribution of angular momenta is $\td P = 2j \td j$~\cite{1919MNRAS..79..408J}. However, numerical studies find that the $j$ distribution for perturbed binaries in the early Universe is $\td P = \td j$~\cite{Raidal:2018bbj}. We will consider distributions
\be
	\frac{\partial P_{p}(j)}{\partial j} = \gamma j^{\gamma-1} 
\ee
with $\gamma \in [1,2]$. Numerical results of Ref.~\cite{Raidal:2018bbj} further imply that
\be\label{eq:K}
	\frac{\partial K(E | E')}{\partial E}
	= \frac{\alpha}{E'}  e^{-\alpha (E/E'-1)} \Theta(E-E') \,,
\ee
 where $\alpha>0$. Binding energies do not change in the limiting case $\alpha\to\infty$, which gives a lower bound on the merger rate. Being perturbed by the initially closest PBH corresponds to $\alpha \approx 1$~\cite{Raidal:2018bbj}. However, binaries may also be perturbed later. Instead of suppressing the merger rate, as in the case of initial binaries, this will lead to an increase in the merger rate because the binaries get harder, i.e. $\alpha$ effectively decreases, while their eccentricity distribution is approximately preserved. The present perturbed binary merger rate can thus be enhanced already when $f_{\rm PBH} \approx 1\%$ as the disruption rate may exceed 50\% in that case (see Fig. \eqref{fig:pnp}).  We neglect binary-binary collisions, which are likely to ionize the wider binary~\cite{Hut:1992wz}.

\begin{figure}
\centering
\includegraphics[width=0.4\textwidth]{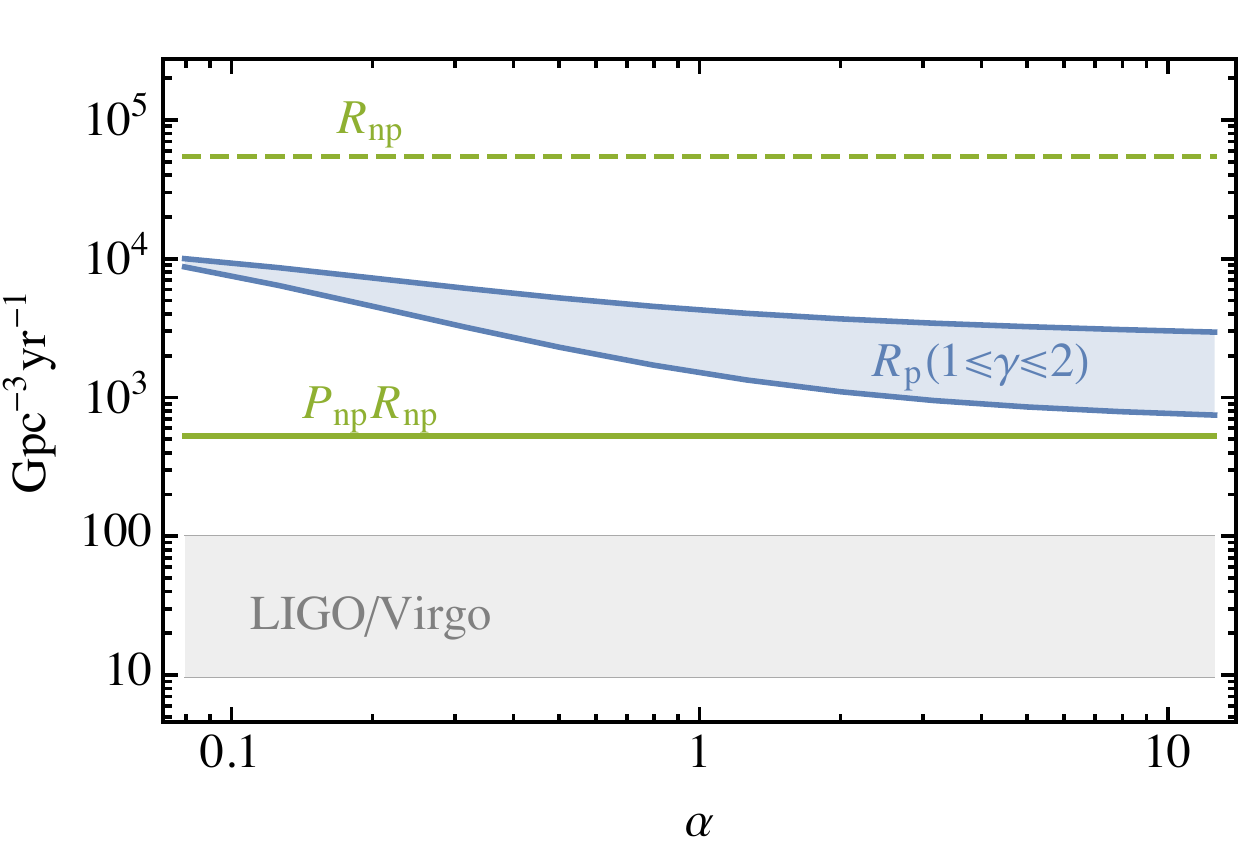}
\includegraphics[width=0.4\textwidth]{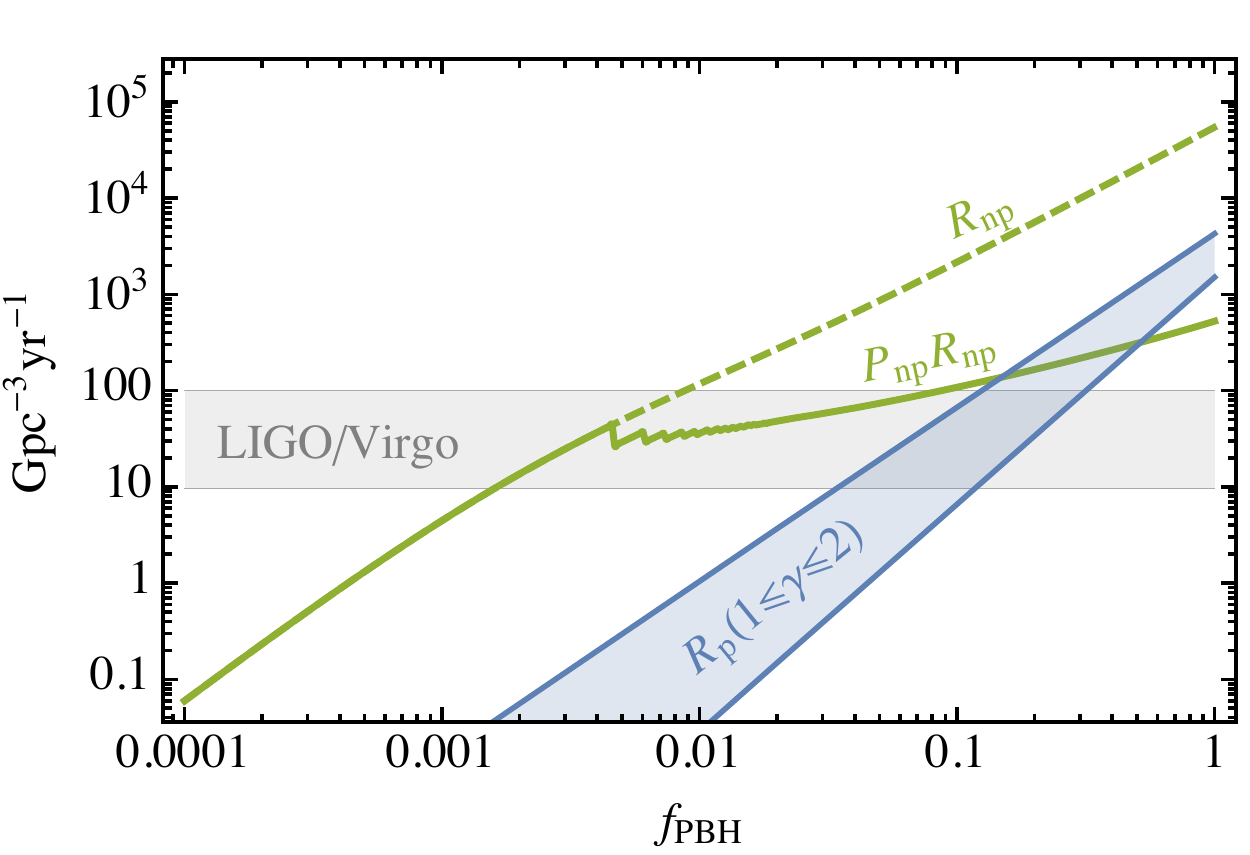}
\caption{The blue lines show the lower bound on the merger rate of perturbed binaries today and the green ones the merger rate of unperturbed initial binaries today for $m=20M_\odot$. The upper bound for the blue region corresponds to $\gamma=1$ and the lower to $\gamma=2$. In the upper panel $f_{\rm PBH}=1$ and in the lower panel $\alpha=1$. The gray band depicts the merger rate indicated by LIGO/Virgo observations. The green solid line and the green dashed line show the merger rate of unperturbed binaries~\eqref{eq:R_unperturbed} with and without the suppression factor~\eqref{eq:pnp}, respectively.}
\label{fig:rates}
\end{figure}

Finally, the merger rate of perturbed binaries at time $t$ is
\bea \label{eq:Rp}
	R_{\rm p}(t) &= \int \td j \, \td E \, \frac{\partial^2 n_{p}(j,E)}{\partial j \partial E} \delta\left(t - \tau(j,E) \right) \\
	&\propto f_{\rm PBH}^{\frac{144\gamma}{259}+\frac{47}{37}} t^{\frac{\gamma}{7}-1} m^\frac{5\gamma-32}{37}  \,.
\eea
When $\gamma>21/37$, this rate grows slower with redshift than the merger rate of unperturbed binaries and scales down faster for decreasing PBH abundance than the rate from unperturbed binaries when $\gamma>7/24$ and $f_{\rm PBH}\gsim 0.01$. It depends weakly on the lower bound of the initial coalescence time given by~\eqref{eq:taubound}, i.e. requiring $\tau_i > q t_p(y)$, we find that $R_{\rm p} \propto q^{\frac{16(28-9\gamma)}{2331}}$.

The present merger rate of perturbed binaries is shown in Fig.~\ref{fig:rates} for $m=20\Msun$. The merger rate decreases as a function of $\alpha$ as seen in the upper panel, and the lower panel shows the dependence of the merger rate on $f_{\rm PBH}$. The merger rate is a decreasing a function of $\gamma$, that is, distributions with more eccentric perturbed binaries correspond to a larger merger rate.  \footnote{In~\cite{Raidal:2017mfl} the rate $R_{\rm p}$ was found to be about 2 orders of magnitude smaller because it was estimated assuming the extremal case where all perturbed binaries have circular orbits. This corresponds to the limit $\gamma \to \infty$.} When $f_{\rm PBH}\approx 1$, the merger rate can be dominated by perturbed binaries. For $f_{\rm PBH} \gtrsim 0.3$ the merger rate $R_{\rm p}$ exceeds the region observed by the LIGO/Virgo collaboration thus ruling out $m=20\Msun$ PBH DM even when all binaries are perturbed.

\section{Discussion}

Consider the modifications due to initial spatial clustering of PBHs~\cite{Belotsky:2018wph}. The effect of a nonvanishing 2-point function $\xi_{\rm PBH}(x)$ on $R_{\rm np}$ in typical models of inflationary PBH production is estimated to be small~\cite{Ballesteros:2018swv}. It should, however, be reconsidered for $R_{\rm p}$. The spatial correlation of PBH can be included by using $N(x) = 4\pi \int \td x x^2 (1+\xi_{\rm PBH}(x))$ in \eqref{eq:npairs}. To find a rough quantitative estimate, we assume a constant 2-point function at scales relevant to binary formation, $1+\xi_{\rm PBH}(x) \approx \delta_{\rm PBH}$ when $N(x) \lesssim 1$~\cite{Raidal:2017mfl}. This is equivalent to a local change in the PBH abundance, so its effect on the merger rate amounts to a simple rescaling 
\be
	R(f_{\rm PBH}) \to \delta_{\rm PBH}^{-1}R(\delta_{\rm PBH} f_{\rm PBH})\,.
\ee
Thus, $R_{\rm p} \propto \delta_{\rm PBH}^{\frac{144\gamma}{259}+\frac{10}{37}}$ is more sensitive to clustering than $R_{\rm np}\propto \delta_{\rm PBH}^{16/37}$. Moreover, clustering will generally lead to more of the initial binaries being perturbed, i.e. a smaller $P_{\rm np}$, and thus to a reduction of $R_{\rm np}$. On the other hand, $R_{\rm p}$ will be enhanced as more frequent 3-body encounters tend to harden the perturbed binaries. Initially clustering therefore enhances $R_{\rm p}$ compared to $R_{\rm np}$.

We expect our results to hold for narrow mass functions. The extension for wider mass functions is nontrivial due to inherent nonlinearities. Mass segregation will enhance the disruption of heavy initial binaries, especially during core collapse, as it forces them to migrate towards the center of the halo. The opposite holds for light binaries, because, although they are more easily perturbed due to their generally lower binding energy, they tend to migrate away from the center, decreasing the probability of hard collisions. Extended mass functions also shorten the collapse timescale~\cite{Meylan:1996yx}, which will make perturbing the binaries more likely. Finally, perturbed binaries are expected to contain PBHs from the heavy end of the mass distribution as the lightest PBH gets ejected in 3-body encounters~\cite{Hut:1992wz}.

\section{A conservative constraint on PBH abundance}

\begin{figure}
\centering
\includegraphics[width=0.4\textwidth]{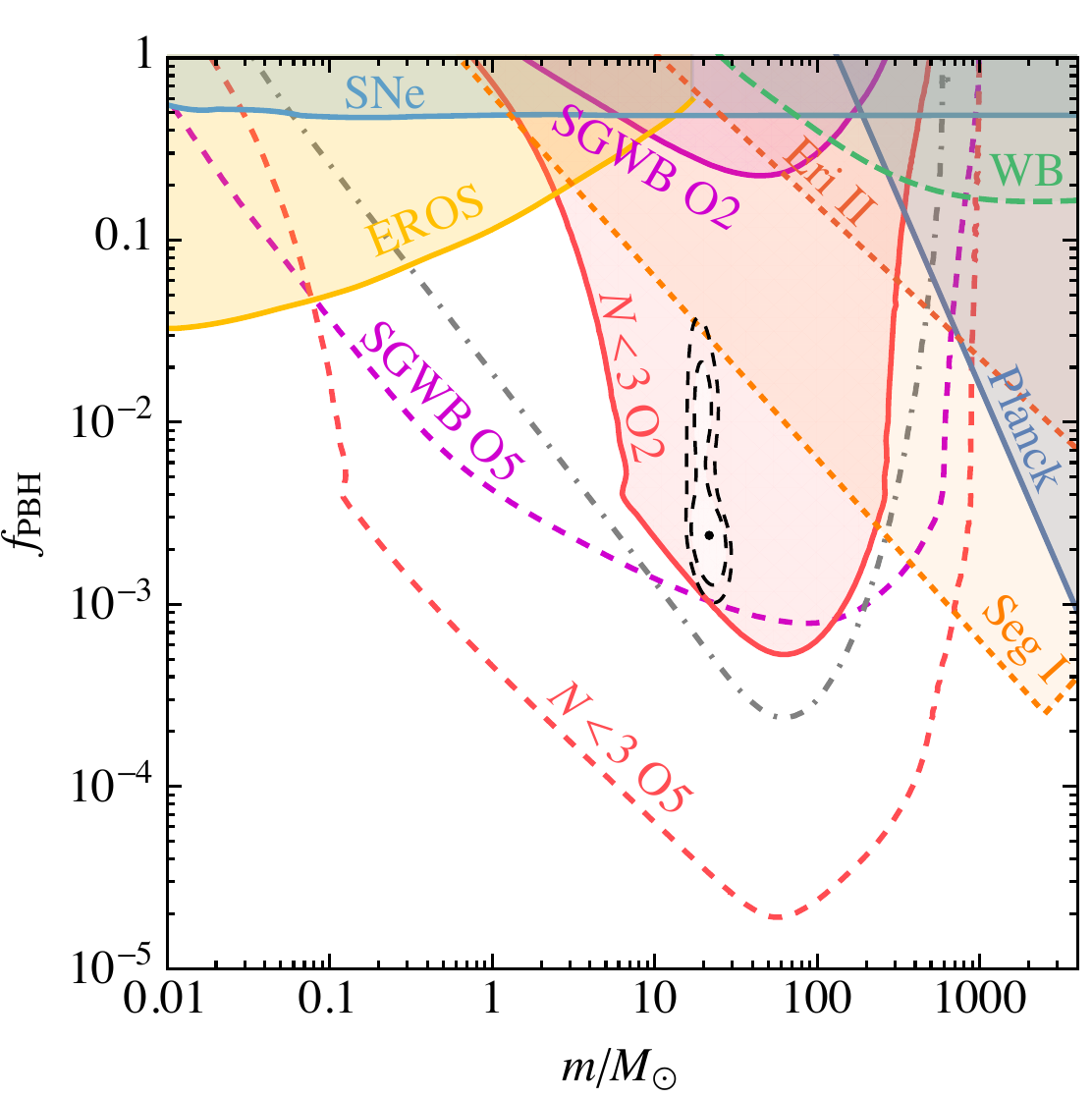}
\caption{Constraints on the PBH abundance for monochromatic mass functions: The red solid line shows the $2\sigma$ constraint from the observed merger rate and the purple solid line the constraint from the nonobservation of the stochastic GW background. The red and purple dashed lines give projections of the final LIGO/Virgo design sensitivity. The gray dot-dashed line shows the $N<3$ O2 constraint if the disruption of binaries, by both mechanisms (1) and (2), is neglected. The black dashed contours depict the likelihood fit for a log-normal PBH mass function with width $\sigma = 0.4$ on the observed rate and masses.}
\label{fig:constraints}
\end{figure}

Constraints arising from the LIGO/Virgo GW measurements for monochromatic PBH mass function are shown in Fig.~\ref{fig:constraints}. For these we used the merger rate $R = P_{\rm np} R_{\rm np} + R_{\rm p}$, where for the parameters of $R_{\rm p}$ we used the values $\alpha=1$ and $\gamma=2$ that yield a smaller rate. Including the rate $R_{\rm p}$ has a relatively small impact on the constraints. The red line shows the $2\sigma$ constraint on the PBH abundance corresponding to the case that none of the BH merger events originated from PBH binaries. This is obtained by calculating the expected number of events~\cite{Raidal:2018bbj},
\be
	N = \mathcal{T} \hspace{-4pt}\int \td R(m_1,m_2,z) \td V_c(z) \theta(\rho(m_1,m_2,z) - \rho_c)\,,
\ee
where $\mathcal{T} = 165$\,days is the observation time for LIGO in observation runs O1 and O2, $\td V_c(z)$ is the differential comoving volume element, $\rho(m_1,m_2,z)^2$ is the signal-to-noise ratio for LIGO observing a merger of BHs of mass $m_1$ and $m_2$ at redshift $z$~\cite{TheLIGOScientific:2016pea}, and $\rho_c = 8$ is the threshold value for detectability of the event~\cite{TheLIGOScientific:2016pea}. The bounds shown by the red lines correspond to $N<3$ which, assuming that the events are Poisson distributed, gives the $95\%$ (or $2\sigma$) upper limit on $f_{\rm PBH}$.

The scenario where all observed BH merger events arise from PBHs takes place in a narrow mass range. For this case we make a maximum likelihood fit of the PBH abundance and mass function (see~\cite{Raidal:2018bbj} for details). The $2\sigma$ and $3\sigma$ contours corresponding to a fit of the observed LIGO/Virgo events to the PBH scenario with a narrow log-normal mass function are shown by the dashed black lines and the best fit by the black dot. 

Faint PBH binary mergers will also contribute to the stochastic GW background. Nonobservation of the latter excludes the region shown in purple. The dashed red and purple lines show the projected final sensitivity of the LIGO/Virgo detectors. The other constraints in this mass range arise from microlensing~\cite{Tisserand:2006zx,Allsman:2000kg,Griest:2013aaa,Zumalacarregui:2017qqd,Garcia-Bellido:2017xvr,Garcia-Bellido:2017imq,Calcino:2018mwh}, dynamics of stars~\cite{Brandt:2016aco,Koushiappas:2017chw,Li:2016utv,Monroy-Rodriguez:2014ula}, and due to accreting PBHs affecting the CMB~\cite{Ricotti:2007au,Horowitz:2016lib,Ali-Haimoud:2016mbv,Poulin:2017bwe} or 21cm physics~\cite{Hektor:2018qqw,Mena:2019nhm,Hutsi:2019hlw}. For comparison, the constraint from the unsuppressed merger rate, given by Eq.~\eqref{eq:R_unperturbed} with $S=1$, is shown by the gray dot-dashed line. 

\section{Conclusion}

In conclusion, we computed the lower bound on the merger rate of PBH binaries by estimating the maximal suppression of the initial PBH binary merger rate. To this aim we assumed that (a) early haloes survive when they are later absorbed by larger structures and (b) haloes small enough to be gravitationally unstable within a the Hubble time will disrupt their binaries with certainty. We remark that omitting assumption (b) would still imply a relatively large suppression as early haloes, due to their larger densities and smaller velocity dispersions, can disrupt their binaries within a  timescale much shorter than the age of the Universe. We further gave a conservative estimate of the merger rate from a population of PBH binaries that were formed in the early Universe from dense 3-body systems and found that the resulting merger rate can exceed the one observed by LIGO/Virgo when $f_{\rm PBH} \gtrsim 0.3$.

These results put the GW constraints on the PBH abundance on a stronger footing. In particular, scenarios where PBHs make up all DM are ruled out in the range $1-100 \Msun$ and LIGO/Virgo design sensitivity has the potential to probe the wide mass range of $10^{-2}- 10^3 \Msun$. We find that the PBH scenarios for the LIGO/Virgo GW events could be realized by relatively narrow PBH mass functions centered around $20\Msun$ with the PBH abundance ranging from $0.1\%-4\%$. Our conclusions persist in models where PBHs are initially clustered as, although clustering makes it more likely for initial binaries to be perturbed, it increases the merger rate of perturbed binaries. As our results are based on analytic models of nonlinear PBH structure formation and binary-PBH interactions, for our conclusions to be considered definitive they should be fully and rigorously tested with numerical simulations, which should predict merger rates higher than found here.

\vspace{1cm}

\acknowledgements We thank Martti Raidal and Yacine Ali-Ha\"{i}moud for useful discussions. This work was supported by the grants IUT23-6, EU through the ERDF CoE program grant TK133,  and by the Estonian Research Council via the Mobilitas Plus grant MOBTP135 and MOBTT5. V.V. was supported by the United Kingdom STFC Grant ST/L000326/1.

\bibliography{PBBH3.bib}

\end{document}